\documentclass[twocolumn,amsmath,showpacs,amsfonts,aps,prc,floatfix]{revtex4}
\usepackage{graphicx}
\usepackage{bm}

\begin{document}

%%\draft
\title{Influence of shear viscosity on the correlation between the triangular flow and initial spatial triangularity}

\author{A. K. Chaudhuri}
\email[E-mail:]{akc@veccal.ernet.in}
\affiliation{Variable Energy Cyclotron Centre, 1/AF, Bidhan Nagar, 
Kolkata 700~064, India}

\begin{abstract}
In a hydrodynamic model, with fluctuating initial conditions, the
correlation between triangular flow and initial spatial triangularity is studied.
The triangular flow, even in ideal fluid, is only weakly correlated with the initial triangularity. The correlation is largely reduced in viscous fluid.
Elliptic flow on the other hand appears to be strongly correlated with initial eccentricity.  Weak correlation between triangular flow and initial triangularity indicate that a   part of triangular flow is unrelated to initial triangularity. Triangularity acquired during the fluid evolution also contributes to the triangular flow.
\end{abstract}


\date{\today}  

\maketitle

In ultra-relativistic nuclear collisions, a deconfined state of quarks and gluons, commonly called   Quark-Gluon-Plasma (QGP) is expected to be produced.  One of the experimental observables of QGP is the azimuthal distribution of the produced particles.  In a non-zero impact parameter collision between two identical nuclei, the collision zone is asymmetric. Multiple collisions transform the initial asymmetry   into momentum anisotropy. Momentum anisotropy is best studied by decomposing it   in a Fourier series, 
 
\begin{equation} \label{eq1}
\frac{dN}{d\phi}=\frac{N}{2\pi}\left [1+ 2\sum_n v_n cos(n\phi-n\psi_n)\right ], n=1,2,3...
\end{equation} 
 
\noindent   $\phi$ is the azimuthal angle of the detected particle and 
$\psi_n$ is the  plane of the symmetry of initial collision zone. For smooth initial matter distribution, plane of symmetry of the collision zone coincides with the reaction plane (the plane containing the impact parameter and the beam axis), 
$\psi_n \equiv \Psi_{RP}, \forall n$. The odd Fourier coefficients are zero by symmetry. However, fluctuations in the positions of the participating nucleons can lead to non-smooth density distribution, which will fluctuate on event-by-event basis.  
The participating nucleons then determine the symmetry plane ($\psi_{PP}$), which fluctuate around the reaction plane \cite{Manly:2005zy}. As a result odd harmonics, which were exactly zero for smoothed initial distribution, can be developed. It has been conjectured that third hadronic $v_3$, which is response of the initial triangularity of the medium, is responsible for the observed structures in two particle correlation in Au+Au collisions \cite{Mishra:2008dm},\cite{Mishra:2007tw},\cite{Takahashi:2009na},\cite{Alver:2010gr},\cite{Alver:2010dn},\cite{Teaney:2010vd}. The ridge structure in pp collisions also has a natural explanation if odd harmonic flow develops.  Recently, ALICE collaboration has observed odd harmonic flows    in Pb+Pb collisions \cite{:2011vk}. In most central collisions, the elliptic flow ($v_2$) and triangular flow ($v_3$) are of similar magnitude. In peripheral collisions however, elliptic flow dominates. 

In a hydrodynamic model, collective flow is a response of the spatial asymmetry of the initial state. For example, elliptic flow   is the response of  ellipticity of the initial medium.  If
ellipticity in the initial medium is characterized by spatial  eccentricity, $\epsilon_2=\frac{<<y^2-x^2>>}{<<y^2+x^2>>}$, more eccentric is the initial medium, more flow will be generated,  $v_2 \propto \epsilon_2$.  
Similar correlation is expected between the triangular flow and initial triangularity. The correlation between triangular flow and initial triangularity is not well studied for fluctuating initial conditions, in particular in viscous fluid. Qualitatively, one can argue that correlation between momentum anisotropy of produced particles and asymmetry of initial density distribution will reduce in presence of viscosity. The reasoning is simple, in ideal hydrodynamics only a single length scale exits in the problem, that of the system size $R$. Viscosity introduces an additional length scale, the microscopic scattering length. The additional length scale will reduce the correlation between flow coefficients and initial spatial asymmetry. In the present brief report, in a hydrodynamic model with fluctuating initial conditions, we have studied the effect of (shear) viscosity on the correlation between triangular flow and initial   triangularity. For comparison, we have also studied the correlation between elliptic flow and initial eccentricity.
In ideal   fluid, elliptic flow shows strong correlation with spatial eccentricity. The correlation is weakened with viscosity. For triangular flow, even in ideal fluid, the correlation is not strong. 
In viscous fluid, it gets even weaker. 

For the fluctuating initial conditions, we have used a model of hot spots in the initial states.  Similar models are used to study   elliptic flow in pp collisions at LHC \cite{CasalderreySolana:2009uk,Bozek:2009dt,Chaudhuri:2009yp}.
 In \cite{Bhalerao:2011bp}, a similar model was used to study anisotropy in heavy ion collisions due to fluctuating initial conditions. 
Recently, we have used the model to study viscous effects on elliptic and triangular flow \cite{arXiv:1108.5552}. 
In the model, it is assumed that  in an impact parameter ${\bf b}$ collision, each participating nucleon pair randomly deposit some energy in the reaction volume, and produces a hot spot. The hot spots are assumed to be Gaussian distributed. The initial energy density is then super position of $N=N_{participant}$ hot spots. 
 
\begin{equation} \label{eq2}
\varepsilon({x,y})=\varepsilon_0 
\sum_{i=1}^{N_{participant}} e^{-\frac{({\bf r}-{\bf r}_i)^2}{2\sigma^2}}
\end{equation}

The participant number $N_{participant}$ is calculated in a Glauber model. We also restrict the centre of hotspots ($r_i$) within the transverse area defined by the Glauber model of participant distribution. The central density $\varepsilon_0$ and the width $\sigma$ are parameters of the model. We fix $\sigma$=1 fm. The central density
$\varepsilon_0$ is fixed to reproduce approximately the experimental charged particles in a peripheral (30-40\%) Pb+Pb collisions. 

We characterize the initial   density distribution in terms of eccentricity $\epsilon_2$ and triangularity $\epsilon_3$.      

\begin{subequations}
\begin{eqnarray}  
\epsilon_2 e^{i 2\psi_2}&=&-\frac{\int \int \varepsilon(x,y) r^2 e^{i 2\phi}dxdy}{\int \int\varepsilon(x,y) r^2 dxdy} \label{eq3a} \\
 \epsilon_3 e^{i 3\psi_3}&=&-\frac{\int \int \varepsilon(x,y) r^3 e^{i 3\phi}dxdy}{\int \int\varepsilon(x,y) r^3 dxdy} \label{eq3b}
\end{eqnarray} 
\end{subequations}

$\psi_2$ and $\psi_3$ in Eq.\ref{eq3a},\ref{eq3b}, are    participant plane angle for elliptic and triangular flow   respectively.
Note that for the triangularity, we have used the definition due to Teaney and Yan
\cite{Teaney:2010vd}. An alternate definition was used by Alver and Rolland \cite{Alver:2010gr}, where $r^3$ terms in Eq.\ref{eq3b} are replaced by $r^2$. However, Teaney and Yan argued from theoretical consideration that in the definition of triangularity, $r^3$ terms are more appropriate that $r^2$.  

Space-time evolution of the fluid was obtained by solving Israel-Stewart's 2nd order theory. We assume that in $\sqrt{s}_{NN}$=2.76 TeV, Pb+Pb collisions  at LHC, a baryon free fluid is formed.
Only dissipative effect we consider is the shear viscosity. Heat conduction and bulk viscosity is neglected. Space-time evolution of the fluid was obtained by solving the following equations, 
 
 \begin{subequations}
\begin{eqnarray}  
\partial_\mu T^{\mu\nu} & = & 0,  \label{eq4a} \\
D\pi^{\mu\nu} & = & -\frac{1}{\tau_\pi} (\pi^{\mu\nu}-2\eta \nabla^{<\mu} u^{\nu>}) \nonumber \\
&-&[u^\mu\pi^{\nu\lambda}+u^\nu\pi^{\mu\lambda}]Du_\lambda. \label{eq4b}
\end{eqnarray}
\end{subequations}

Eq.\ref{eq4a} is the conservation equation for the energy-momentum tensor, $T^{\mu\nu}=(\varepsilon+p)u^\mu u^\nu - pg^{\mu\nu}+\pi^{\mu\nu}$, 
$\varepsilon$, $p$ and $u$ being the energy density, pressure and fluid velocity respectively. $\pi^{\mu\nu}$ is the shear stress tensor. Eq.\ref{eq4b} is the relaxation equation for the shear stress tensor $\pi^{\mu\nu}$.   
In Eq.\ref{eq4b}, $D=u^\mu \partial_\mu$ is the convective time derivative, $\nabla^{<\mu} u^{\nu>}= \frac{1}{2}(\nabla^\mu u^\nu + \nabla^\nu u^\mu)-\frac{1}{3}  
(\partial . u) (g^{\mu\nu}-u^\mu u^\nu)$ is a symmetric traceless tensor. $\eta$ is the shear viscosity and $\tau_\pi$ is the relaxation time.  It may be mentioned that in a conformally symmetric fluid the relaxation equation can contain additional terms  \cite{Song:2008si}. Assuming boost-invariance, the equations are solved  in $(\tau=\sqrt{t^2-z^2},x,y,\eta_s=\frac{1}{2}\ln\frac{t+z}{t-z})$ coordinates, with the code 
  "`AZHYDRO-KOLKATA"', developed at the Cyclotron Centre, Kolkata.
 Details of the code can be found in \cite{Chaudhuri:2008sj}.
  
Hydrodynamic equations are closed with an equation of state (EoS) $p=p(\varepsilon)$.
Currently, there is consensus that the confinement-deconfinement transition is a cross over and the cross over or the pseudo critical temperature for the  transition  is
$T_c\approx$170 MeV \cite{Aoki:2006we,Aoki:2009sc,Borsanyi:2010cj,Fodor:2010zz}.
In the present study, we use an equation of state where the Wuppertal-Budapest \cite{Aoki:2006we,Borsanyi:2010cj} 
lattice simulations for the deconfined phase is smoothly joined at $T=T_c=174$ MeV, with hadronic resonance gas EoS comprising all the resonances below mass $m_{res}$=2.5 GeV. Details of the EoS can be found in \cite{Roy:2011xt}.
  
In addition to the initial energy density for which we use the model of hot spots,  solution of partial differential equations (Eqs.\ref{eq4a},\ref{eq4b}) requires to specify the fluid velocity ($v_x(x,y),v_y(x,y)$) and shear stress tensor ($\pi^{\mu\nu}(x,y)$) at the initial time $\tau_i$. One also need to specify the viscosity ($\eta$) and the relaxation time ($\tau_\pi$). A freeze-out prescription is also needed to convert the information about fluid energy density and velocity to particle spectra.  We assume that the fluid is thermalized at $\tau_i$=0.6 fm and the initial fluid velocity is zero, $v_x(x,y)=v_y(x,y)=0$. We initialize the
shear stress tensor  to boost-invariant values, $\pi^{xx}=\pi^{yy}=2\eta/3\tau_i$, $\pi^{xy}$=0 and for  the relaxation time, we use the   Boltzmann estimate $\tau_\pi=3\eta/2p$. We also assume that the viscosity to entropy density ($\eta/s$) remains a constant throughout the evolution and simulate Pb+Pb collisions for a range of $\eta/s$. The freeze-out is fixed at $T_F$=130 MeV.

  \begin{figure}[t]
 %\vspace{0.3cm} 
 \center
 \resizebox{0.30\textwidth}{!}{%
  \includegraphics{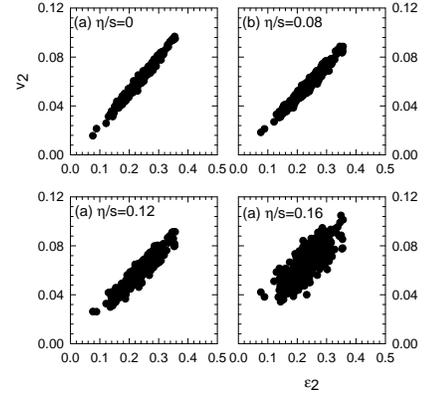} 
}
\caption{In four panels (a)-(d), for fluid viscosity to entropy ratio $\eta/s$=0, 0.08, 0.12 and 0.16, the correlation between   elliptic flow ($v_2$) and initial eccentricity ($\epsilon_2$) is shown. Event size is $N_{event}$=500.}
\label{F1}
\end{figure} 
 
For fluid viscosity to entropy ratio $\eta/s$=0, 0.08, 0.12 and 0.16, we have simulated b=8.9 fm Pb+Pb collisions. b=8.9 fm collisions approximately corresponds to 30-40\% collision. In viscous evolution, entropy is generated. To account for the entropy generation, the Gaussian density $\varepsilon_0$ was reduced with increasing viscosity, such that in ideal and viscous fluid, on the average, $\pi^-$ multiplicity remains the same. In the present study, we have used  $N_{event}$=500 events.
In each event, Israel-Stewart's  hydrodynamic equations are solved and from the freeze-out surface, invariant distribution ($\frac{dN}{dyd^2p_T}$) for $\pi^-$ was obtained. 
In analogy to Eq.\ref{eq3a},\ref{eq3b}, invariant distribution can be characterized by 'harmonic flow coefficients' \cite{arXiv:1104.0650}.

\begin{eqnarray}
%v_n(y,p_T)e^{i\psi_n(y,p_T)}&=&\frac{\int d\phi e^{in\phi} \frac{dN}{d\phi p_Tdp_T}}  {\frac{dN}{d\phi p_Tdp_T}}\\
 v(y)_ne^{i\psi_n(y)}&=&\frac{\int p_T dp_T d\phi e^{in\phi} \frac{dN}{d\phi p_Tdp_T}} 
 {\frac{dN}{d\phi}}, n=2,3
\end{eqnarray}

 \begin{figure}[t]
 %\vspace{0.3cm} 
 \center
 \resizebox{0.30\textwidth}{!}{%
  \includegraphics{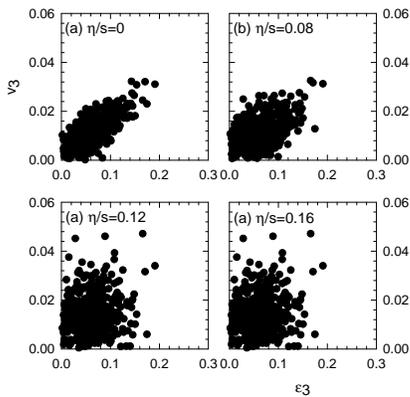} 
}
\caption{same as in Fig.\ref{F1}, but for the  triangular flow($v_3$) and initial triangularity ($\epsilon_3$).}
\label{F2}
\end{figure}

In a boost-invariant version of hydrodynamics, flow coefficients are rapidity independent and in the following, we drop the rapidity dependence. 
Present simulations are applicable only in the central rapidity region, $y\approx$0, where boost-invariance is most justified. 

 \begin{table}[ht]%%{\bf Glauber} %%\\[1ex]
\caption{\label{table1} Correlation measure for elliptic and triangular flow, as a function of viscosity over entropy ratio.} 
\begin{ruledtabular} 
  \begin{tabular}{|c|c|c|c|c|}\hline
  & $\frac{\eta}{s}$=0 &  $\frac{\eta}{s}$=0.08 & $\frac{\eta}{s}$=0.12  & $\frac{\eta}{s}$=0.16\\\hline
 $C_{measure}(v_2)$ & $\frac{0.46}{8.77}$   & $\frac{0.53}{8.77}$ & $\frac{0.92}{8.77}$  & $\frac{1.77}{8.77}$\\\hline
 $C_{measure}(v_3)$ & $\frac{2.46}{8.77}$   & $\frac{2.95}{8.77}$ & $\frac{3.91}{8.77}$  & $\frac{4.50}{8.77}$\\\hline
  \end{tabular}\end{ruledtabular}  
\end{table}  

In Fig.\ref{F1}, in 4 panels, for fluid viscosity $\eta/s$=0, 0.08, 0.12 and 0.16, simulated elliptic flow ($v_2$) is plotted against initial eccentricity ($\epsilon_2$). Each panel contains 500 data points. In ideal fluid, $v_2$ and $\epsilon_2$ are strongly correlated, $v_2 \propto \epsilon_2$. Evidently, the correlation is gradually weakened as the viscosity of the fluid is increased. The result is not unexpected. As argued earlier, in viscous fluid, correlation between elliptic flow and initial eccentricity is reduced due to introduction of the additional length scale.    Correlation between triangular flow ($v_3$) and initial triangularity ($\epsilon_3$) is studied in Fig.\ref{F2}. The results are more interesting. Even in ideal fluid, $v_3$ and $\epsilon_3$ are not strongly correlated. In viscous fluid correlations are even worse. 
Indeed, for fluid viscosity $\eta/s$=0.12-0.16, it appears that flow coefficients are marginally related to initial triangularity.

Qualitatively,  from Fig.\ref{F1} and \ref{F2}, one understand that the correlation between triangular flow and initial triangularity is   much less  than the correlation between the elliptic flow and initial eccentricity. One also understands that the correlation reduces with viscosity. We can  obtain  a quantitative measure of   the correlation between flow coefficients ($v_{2/3}$) and initial asymmetry parameter ($\epsilon_{2/3}$). We note that for a perfect correlation, $v_n \propto \epsilon_n$ and simulated flow coefficients will fall on a straight line. Dispersion of the flow coefficients around the best fitted straight line then gives a measure of the correlation. 
We thus define a correlation measure function $C_{measure}$,

\begin{equation} \label{eq6}
C_{measure}(v_n)=\frac {\sum_i[v^i_{n,sim}(\epsilon_n)-v^i_{n,st.line}(\epsilon_n)]^2}
{\sum_i[V^i_{random}(\epsilon)-V^i_{st.line}(\epsilon)]^2},
\end{equation}

\noindent which measure the dispersion of the simulated flow coefficients from a best fitted straight line, relative to   completely random flow coefficients \cite{note}. In order to compare the correlation between $v_2$ and $\epsilon_2$ and the correlation between $v_3$ and $\epsilon_3$,  we also rescale the flow coefficients ($v_{2/3}$) and asymmetry parameters ($\epsilon_{2/3}$) to  vary between (0-1), such that the dispersion is measured in a common scale. $C_{measure}$ varies between 0-1.
If simulated flows  are perfectly correlated with the asymmetry measure,   $C_{measure}=0$,  at the opposite limit, when they are perfectly uncorrelated (random) $C_{measure}=1$. 
In table.\ref{table1}, we have noted $C_{measure}$ for elliptic and triangular flow, as a function of $\eta/s$. With increasing fluid viscosity, correlation between the flow coefficient and spatial asymmetry parameter is reduced. In ideal  and minimally viscous fluid, elliptic flow and initial eccentricity are mostly correlated, $C_{measure}(v_2)\approx 0.05$. For more viscous fluid  $\eta/s$=0.12-0.16, correlation though reduced, remains strong, $C_{measure}(v_2)\approx 0.1-0.2$.  If we interpret $C_{measure}$ as the fraction of flow unrelated to the initial spatial asymmetry, for elliptic flow, the fraction is small, less than  $\sim$10\% for fluid viscosity over entropy ratio $\eta/s$=0-0.12. 
Triangular flow on the other hand appears to be highly uncorrelated. Even for ideal fluid, $C_{measure}(v_3)\approx 0.3$. In the above interpretation, $\sim$30\% of triangular flow is unrelated to the initial triangularity. For more viscous fluid, $\eta/s$=0.12-0.16, the fraction is increased to $\sim$50\%. 
Comparatively large value of $C_{measure}$ for triangular flow raises an important question. Is triangular flow is response of the initial triangularity of the medium only?  
One may argue that in RHIC/LHC energy collisions, viscosity of  the produced fluid is not large and the simulation results for fluid viscosity $\eta/s$=0.12-0.16 will not be of any practical concern. However the correlation between $v_3$ and $\epsilon_3$ is still weak in ideal or minimally viscous fluid.
While for elliptic flow, $\sim$95\% of the flow is related to initial spatial asymmetry,  for the triangular flow, the fraction is only $\sim$65-70\%.
It is reasonable to conjecture that a large part of the triangular flow is unrelated to the initial triangularity.   What are the mechanisms by which the system acquires triangularity is uncertain. Processes like jet quenching, Cerenkov radiation etc. may introduce triangularity in the system. Indeed, hydrodynamical simulations of jet quenching do indicate   development of  triangularity in density distribution at late time  (see Fig.2 of ref.\cite{Chaudhuri:2006qk}).

To summarize, in a hydrodynamic model, with fluctuating initial conditions, we have studied the correlation between triangular flow and initial triangularity of the medium. In ideal or minimally viscous fluid, triangular flow is only weakly correlated with initial triangularity. In more viscous fluid, the correlation gets even weaker. Elliptic flow on the other hand is strongly correlated with initial eccentricity in ideal or   viscous fluid. 
  Weak correlation between triangular flow and initial triangularity strongly indicate that a part of the triangular flow   is  unrelated to initial triangularity of the medium. Final state triangularity, generated by unknown mechanisms, also contributes to the triangular flow.


\begin{thebibliography}{99}
%\cite{Manly:2005zy}
\bibitem{Manly:2005zy}
  S.~Manly {\it et al.}  [PHOBOS Collaboration],
  %``System size, energy and pseudorapidity dependence of directed and elliptic
  %flow at RHIC,''
  Nucl.\ Phys.\  A {\bf 774}, 523 (2006)
%  [arXiv:nucl-ex/0510031].
  %%CITATION = NUPHA,A774,523;%%
%\cite{Mishra:2008dm}
\bibitem{Mishra:2008dm}
  A.~P.~Mishra, R.~K.~Mohapatra, P.~S.~Saumia, A.~M.~Srivastava,
  %``Using cosmic microwave background radiation analysis tools for flow anisotropies in relativistic heavy-ion collisions,''
  Phys.\ Rev.\  {\bf C81}, 034903 (2010).
%  [arXiv:0811.0292 [hep-ph]].
%\cite{Mishra:2007tw}
\bibitem{Mishra:2007tw}
  A.~P.~Mishra, R.~K.~Mohapatra, P.~S.~Saumia, A.~M.~Srivastava,
  %``Super-horizon fluctuations and acoustic oscillations in relativistic heavy-ion collisions,''
  Phys.\ Rev.\  {\bf C77}, 064902 (2008).
%  [arXiv:0711.1323 [hep-ph]].
%\cite{Takahashi:2009na}
\bibitem{Takahashi:2009na}
  J.~Takahashi, B.~M.~Tavares, W.~L.~Qian, R.~Andrade, F.~Grassi, Y.~Hama, T.~Kodama, N.~Xu,
  %``Topology studies of hydrodynamics using two particle correlation analysis,''
  Phys.\ Rev.\ Lett.\  {\bf 103}, 242301 (2009).
 % [arXiv:0902.4870 [nucl-th]].
%\cite{Alver:2010gr}
\bibitem{Alver:2010gr}
  B.~Alver, G.~Roland,
  %``Collision geometry fluctuations and triangular flow in heavy-ion collisions,''
  Phys.\ Rev.\  {\bf C81}, 054905 (2010).
 % [arXiv:1003.0194 [nucl-th]].
%\cite{Alver:2010dn}
\bibitem{Alver:2010dn}
  B.~H.~Alver, C.~Gombeaud, M.~Luzum, J.~-Y.~Ollitrault,
  %``Triangular flow in hydrodynamics and transport theory,''
  Phys.\ Rev.\  {\bf C82}, 034913 (2010).
 % [arXiv:1007.5469 [nucl-th]].
%\cite{Teaney:2010vd}
\bibitem{Teaney:2010vd}
  D.~Teaney, L.~Yan,
  %``Triangularity and Dipole Asymmetry in Heavy Ion Collisions,''
  Phys.\ Rev.\  {\bf C83}, 064904 (2011).
%  [arXiv:1010.1876 [nucl-th]].
%\cite{:2011vk}
\bibitem{:2011vk}
  [ ALICE Collaboration ],
  %``Higher harmonic anisotropic flow measurements of charged particles in Pb-Pb collisions at sqrt(s_{(NN)}) = 2.76 TeV,''
  Phys.\ Rev.\ Lett.\  {\bf 107}, 032301 (2011).
 % [arXiv:1105.3865 [nucl-ex]].
%\cite{CasalderreySolana:2009uk}
\bibitem{CasalderreySolana:2009uk}
  J.~Casalderrey-Solana, U.~A.~Wiedemann,
  %``Eccentricity fluctuations make flow measurable in high multiplicity p-p collisions,''
  Phys.\ Rev.\ Lett.\  {\bf 104}, 102301 (2010).
  [arXiv:0911.4400 [hep-ph]].



%\cite{Bozek:2009dt}
\bibitem{Bozek:2009dt}
  P.~Bozek,
  %``Observation of the collective flow in proton-proton collisions,''
  Acta Phys.\ Polon.\  {\bf B41}, 837 (2010).
 % [arXiv:0911.2392 [nucl-th]].



   
    %\cite{Chaudhuri:2009yp}
\bibitem{Chaudhuri:2009yp}
  A.~K.~Chaudhuri,
  %``Large elliptic flow in low multiplicity pp collisions at LHC energy
  %$\sqrt{s}$=14 TeV,''
  Phys.\ Lett.\  B {\bf 692}, 15 (2010)
 % [arXiv:0912.2578 [nucl-th]].
  %%CITATION = PHLTA,B692,15;%% 

%\cite{Bhalerao:2011bp}
\bibitem{Bhalerao:2011bp}
  R.~S.~Bhalerao, M.~Luzum, J.~-Y.~Ollitrault,
  %``Understanding anisotropy generated by fluctuations in heavy-ion collisions,''
 [arXiv:1107.5485 [nucl-th]].

%\cite{arXiv:1108.5552}
\bibitem{arXiv:1108.5552} 
  A.~K.~Chaudhuri,
  %``Fluctuating initial conditions and fluctuations in elliptic and triangular flow,''
  arXiv:1108.5552 [nucl-th].
  %%CITATION = ARXIV:1108.5552;%% 

 
 
 
   %\cite{Song:2008si}
\bibitem{Song:2008si}
  H.~Song and U.~W.~Heinz,
  %``Multiplicity scaling in ideal and viscous hydrodynamics,''
  Phys.\ Rev.\  C {\bf 78}, 024902 (2008).%[arXiv:0805.1756 [nucl-th]].
  %%CITATION = PHRVA,C78,024902;%%
  
 %\cite{Chaudhuri:2008sj}
\bibitem{Chaudhuri:2008sj} A.~K.~Chaudhuri,
  %``Viscous fluid dynamics in Au+Au collisions at RHIC,''
 arXiv:0801.3180 [nucl-th].
  %%CITATION = ARXIV:0801.3180;%%      
    %\cite{Aoki:2006we}
\bibitem{Aoki:2006we}
  Y.~Aoki, G.~Endrodi, Z.~Fodor, S.~D.~Katz and K.~K.~Szabo,
  %``The order of the quantum chromodynamics transition predicted by the
  %standard model of particle physics,''
  Nature {\bf 443}, 675 (2006)
%  [arXiv:hep-lat/0611014].
  %%CITATION = NATUA,443,675;%%
%\cite{Aoki:2009sc}
\bibitem{Aoki:2009sc}
  Y.~Aoki, S.~Borsanyi, S.~Durr, Z.~Fodor, S.~D.~Katz, S.~Krieg and K.~K.~Szabo,
  %``The QCD transition temperature: results with physical masses in the
  %continuum limit II,''
  JHEP {\bf 0906}, 088 (2009)
%  [arXiv:0903.4155 [hep-lat]].
  %%CITATION = JHEPA,0906,088;%%
 %\cite{Fodor:2010zz}
 %\cite{Borsanyi:2010cj}
\bibitem{Borsanyi:2010cj}
  S.~Borsanyi {\it et al.},
  %``The QCD equation of state with dynamical quarks,''
  JHEP {\bf 1011}, 077 (2010)
  [arXiv:1007.2580 [hep-lat]].
  %%CITATION = JHEPA,1011,077;%%
 
 \bibitem{Fodor:2010zz}
  Z.~Fodor,
  %``QCD thermodynamics on the lattice: Approaching the continuum limit with
  %physical quark masses,''
  J.\ Phys.\ Conf.\ Ser.\  {\bf 230} (2010) 012013.
  %%CITATION = 00462,230,012013;%%
  
%\cite{Roy:2011xt}
\bibitem{Roy:2011xt}
  V.~Roy and A.~K.~Chaudhuri, Phys. Lett. B (in press)
  %``Charged particle's elliptic flow in 2+1D viscous hydrodynamics at LHC
  %($\sqrt{s}$= 2.76 TeV) energy in Pb+Pb collision,''
  arXiv:1103.2870 [nucl-th].
  %%CITATION = ARXIV:1103.2870;%%  
%\cite{arXiv:1104.0650}
\bibitem{arXiv:1104.0650} 
  Z.~Qiu and U.~W.~Heinz,
  %``Event-by-event shape and flow fluctuations of relativistic heavy-ion collision fireballs,''
  Phys.\ Rev.\ C\ {\bf 84}, 024911  (2011)
  [arXiv:1104.0650 [nucl-th]].
  %%CITATION = PHRVA,C84,024911;%%

\bibitem{note} the denominator of Eq.\ref{eq6} is computed by generating   500 random $v_2$ and $\epsilon_2$ and  explicity computed the dispersion from a best fitted striaght line.

%\cite{Chaudhuri:2006qk}
\bibitem{Chaudhuri:2006qk}
  A.~K.~Chaudhuri,
  %``Conical flow due to partonic jets in central Au + Au collisions,''
  Phys.\ Rev.\  C {\bf 75}, 057902 (2007).
  %[arXiv:nucl-th/0610121].
  %%CITATION = PHRVA,C75,057902;%%
  
  
\end{thebibliography}
\end{document}